\documentclass[12pt]{article}

\usepackage{graphicx}

\def\gsim{\raise0.3ex\hbox{$\;>$\kern-0.75em\raise-1.1ex\hbox{$\sim\;$}}}
\def\lsim{\raise0.3ex\hbox{$\;<$\kern-0.75em\raise-1.1ex\hbox{$\sim\;$}}}

\newcommand{\AddrAHEP}{
  {\it AHEP Group, Instituto de F\'{\i}sica Corpuscular --
    C.S.I.C./Universitat de Val{\`e}ncia \\
    Edificio de Institutos de Paterna, Apartado 22085,
  E--46071 Val{\`e}ncia, Spain}}

\begin{document}

\begin{titlepage}
  
\begin{flushright}
hep-ph/0609307 \\
IFIC/06-31\\
\end{flushright}
\vspace*{3mm}  

\begin{center}
\textbf{{\large Experimental tests for the Babu-Zee two--loop model of 
Majorana neutrino masses}}
\\[10mm]
D. Aristizabal Sierra and M. Hirsch 
\vspace{0.3cm}\\
\AddrAHEP.\vspace{0.3cm}\\
\end{center}

\begin{abstract}
The smallness of the observed neutrino masses might have a radiative 
origin. Here we revisit a specific two-loop model of neutrino mass, 
independently proposed by Babu and Zee. We point out that current 
constraints from neutrino data can be used to derive strict lower 
limits on the branching ratio of flavour changing charged lepton 
decays, such as $\mu \rightarrow e \gamma$. Non-observation of 
Br($\mu \rightarrow e \gamma$) at the level of $10^{-13}$ would 
rule out singly charged scalar masses smaller than 590 GeV (5.04 TeV) 
in case of normal (inverse) neutrino mass hierarchy. Conversely, decay 
branching ratios of the non-standard scalars of the model can be fixed 
by the measured neutrino angles (and mass scale). Thus, if the scalars 
of the model are light enough to be produced at the LHC or ILC, measuring 
their decay properties would serve as a direct test of the model as the 
origin of neutrino masses. 
\end{abstract}
\end{titlepage}

\section{Introduction}

During the past few years neutrino oscillation experiments have firmly 
established that neutrinos have non-zero masses and mixing angles among 
the different generations \cite{Fukuda:1998mi}. While for the absolute 
scale of neutrino mass only upper limits of the order 
$m_{\nu}\sim {\cal O}(2 \hskip1mm {\rm eV})$  
exist \cite{PDG}, two neutrino mass squared differences and two neutrino 
angles are by now known quite precisely \cite{Maltoni:2004ei}. These are 
the atmospheric neutrino mass, $\Delta m^2_{\rm Atm} = (2.0-3.2)$ [$10^{-3}$ 
eV$^2$], and angle, $\sin^2\theta_{\rm Atm} = (0.34-0.68)$, as well as the 
solar neutrino mass $\Delta m^2_{\odot} = (7.1-8.9)$ [$10^{-5}$ eV$^2$], 
and angle, $\sin^2\theta_{\odot} = (0.24-0.40)$, all numbers at 3 $\sigma$ 
c.l. For the remaining neutrino angle, the so-called Chooz 
\cite{Apollonio:2002gd} or reactor neutrino angle $\theta_R$, a global fit 
to all neutrino data \cite{Maltoni:2004ei} currently gives a limit of 
$\sin^2\theta_R \le 0.04$ $@$ 3 $\sigma$ c.l.

From a theoretical perspective, there exist several options to explain 
the smallness of the observed neutrino masses. Perhaps the simplest - 
but certainly the most popular - possibility is the seesaw mechanism 
\cite{Minkowski:1977sc,seesaw}. Many variants of the seesaw exist, 
see for example the recent review \cite{Valle:2006vb}. However, most 
realizations of the seesaw make use of a large scale, typically 
the Grand Unification Scale, to suppress neutrino masses and are, 
therefore, only indirectly testable. 

On the other hand, many neutrino mass models exist, in which the scale of 
lepton number violation can be as low as the electro-weak scale or lower. 
Examples are supersymmetric models with violation of R-parity 
\cite{HallSuzuki,hirsch:2000ef}, models with Higgs triplets 
\cite{Schechter:1980gr} or a combination of both 
\cite{AristizabalSierra:2003ix}, leptoquarks \cite{Nieves:1981tv} or 
radiative models, both with neutrino masses at 1-loop 
\cite{Zee:1980ai,AristizabalSierra:2006ri} or at 2-loop 
\cite{Nieves:1981tv,Zee:1985id,Babu:1988ki} order. Radiative mechanisms 
might be considered especially appealing, since they generate small 
neutrino masses automatically, essentially due to loop factors. 

\begin{figure}
\begin{center}
\includegraphics[width=60mm, height=30mm]{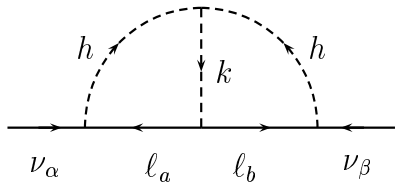}
\end{center}
\caption{Feynman diagram for the 2-loop Majorana neutrino masses in 
the model of \cite{Zee:1985id,Babu:1988ki}.}
\label{fig:Mnu}
\end{figure}

In this paper we will concentrate on a model of neutrino masses, proposed 
independently by Zee \cite{Zee:1985id} and Babu \cite{Babu:1988ki}, in 
which neutrino masses arise only at 2-loop order. The model introduces 
two new charged scalars, $h^+$ and $k^{++}$, both singlets under $SU(2)_L$, 
which couple only to leptons. One can easily estimate, see fig. 
(\ref{fig:Mnu}) and the discussion in the next section, that neutrino masses 
in this setup are of order 
$m_{\nu} \sim (f^2 h)/(16 \pi^2)^2 (m_{\mu}^2/m_{S})$, 
i.e. ${\cal O} \sim 1$ eV for couplings $f$ and $h$ of order ${\cal O}(1)$ 
and scalar mass parameters, $m_S$, of order ${\cal O}(100)$ GeV. Given that 
current neutrino data requires at least one neutrino to have a mass of order 
${\cal O}(0.05)$ eV, one expects that the new scalars should have masses in 
the range ${\cal O}(0.1-1)$ TeV. The model is therefore potentially 
testable at near-future accelerators, such as the LHC or ILC.

Babu and Macesanu \cite{Babu:2002uu} recently re-analyzed this model in 
light of solar and atmospheric neutrino oscillation data. They identified 
the regions in parameter space, in which the model can explain the 
experimental neutrino data and tabulated in some detail constraints on 
the model parameters, which can be derived from the non-observation of 
various lepton flavour violating decay processes. Here, we extend upon 
the results presented in \cite{Babu:2002uu} by pointing out that 
(a) current neutrino data can be used to derive absolute lower limits 
on the branching ratios of the processes $l_{\alpha}\rightarrow l_{\beta}
\gamma$. Especially important in view of future experimental sensitivities 
\cite{exp:meg} is that $Br(\mu\rightarrow e\gamma)\ge 10^{-13}$ is guaranteed 
for charged scalar masses smaller than 590 GeV (5.04 TeV) in case of normal 
(inverse) neutrino mass hierarchy. And (b) decay 
branching ratios of the non-standard scalars of the model can be fixed 
by the measured neutrino angles (and mass scale). Thus, if the scalars 
of the model are light enough to be produced at the LHC or ILC, measuring 
their decay properties would serve as a direct test of the model as the 
origin of neutrino masses. 

The rest of this paper is organized as follows. In the next section, we 
discuss the Lagrangian of the model, as well as its parameters in light 
of current oscillation data. In this part we will make extensive use of the 
results of \cite{Babu:2002uu}. In section 3, we calculate flavour violating 
charged lepton decays, $l_{a} \rightarrow l_{b}l_{c}l_{d}$ and 
$l_{\alpha}\rightarrow l_{\beta}\gamma$, discussing their connection with 
neutrino physics in some detail. Then, we consider the decays of the new 
scalars at future colliders, presenting ranges for various decay branching 
ratios as predicted by current neutrino data. We then close with a short 
discussion.

\section{Neutrino masses at 2-loop}

As mentioned above, the model we consider \cite{Zee:1985id,Babu:1988ki} 
is a simple extension of the standard model, containing two new scalars, 
$h^+$ and $k^{++}$, both singlets under $SU(2)_L$. Their coupling to 
standard model leptons is given by
\begin{equation}\label{yuks}
{\cal L} = f_{\alpha\beta} (L^{Ti}_{\alpha L}CL^{j}_{\beta L})\epsilon_{ij}h^+
         + h'_{\alpha\beta}(e^T_{\alpha R}Ce_{\beta R})k^{++} + {\rm h.c.}
\end{equation}
Here, $L_L$ are the standard model (left-handed) lepton doublets, $e_R$ 
the charged lepton singlets, $\alpha ,\beta$ are generation indices and 
$\epsilon_{ij}$ is the completely antisymmetric tensor. Note that $f$ 
is antisymmetric, while $h'$ is symmetric. Assigning $L=2$ to $h^-$ and 
$k^{++}$, eq.  (\ref{yuks}) conserves lepton number. Lepton number violation 
in the model resides only in the following term in the scalar potential
\begin{equation}\label{scalar}
{\cal L} = - \mu h^+h^+k^{--} + {\rm h.c.}
\end{equation}
Here, $\mu$ is a parameter with dimension of mass, its value is not 
predicted by the model. However, vacuum stability arguments can be used 
to derive an {\em upper bound} for this parameter \cite{Babu:2002uu}. 
For $m_{h}\sim m_{k}$ this bound reads 
\begin{equation}\label{limmu}
\mu \le (6\pi^2)^{1/4}m_h. 
\end{equation}
The setup of eq. (\ref{yuks}) and eq. (\ref{scalar}) generates Majorana 
neutrino masses via the two-loop diagram shown in fig. (\ref{fig:Mnu}). 
The resulting neutrino mass matrix can be expressed as 
\begin{equation}\label{mnu}
{\cal M}^{\nu}_{\alpha\beta} = \frac{8\mu}{(16 \pi^2)^2 m_h^2}
f_{\alpha x}\omega_{xy}f_{y\beta}{\cal I}(\frac{m_{k}^2}{m_{h}^2}), 
\end{equation}
with summation over $x,y$ implied. The parameters $\omega_{xy}$ are defined 
as $\omega_{xy}= m_x h_{xy} m_y$, with $m_x$ the mass of the charged lepton 
$l_x$. Following \cite{Babu:2002uu} we have rewritten $h_{\alpha\alpha}= 
h'_{\alpha\alpha}$ and $h_{\alpha\beta}=2 h'_{\alpha\beta}$. ${\cal I}(r)$ 
finally is a dimensionless two-loop integral given by 
\footnote{We correct a minor misprint in eq.(7) of \cite{Babu:2002uu}.}
\begin{equation}\label{itilde}
{\cal I}(r) = - \int_0^1 dx \int_0^{1-x} dy 
\frac{1}{x+(r-1)y+y^2}\log\frac{y(1-y)}{x+ry} .
\end{equation}
For non-zero values of $r$, ${\cal I}(r)$ can be solved only numerically. 
We note that for the range of interest, say $10^{-2} \le r \le 10^{2}$, 
${\cal I}(r)$ varies quite smoothly between (roughly) $3 \le {\cal I}(r) \le 
0.2$. 

Eq.(\ref{mnu}) generates only two non-zero neutrino masses. This can easily 
be seen from its index structure: ${\rm Det}({\cal M}^{\nu}) = 
{\rm Det}(f_{\alpha x}\omega_{xy}f_{y\beta}) = {\rm Det}(f_{\alpha\beta})=0$. 
The model therefore can not generate a degenerate neutrino spectrum. One 
can find the eigenvector for the massless state, it is proportional 
to
\begin{equation}\label{v0t}
v_0^T = {\cal N}(1,-\epsilon,\epsilon')
\end{equation}
where ${\cal N} = (1 +\epsilon^2+\epsilon'^2)^{-1/2}$ is a 
normalization factor. Here we have introduced  
\begin{eqnarray}\label{defeps}
\epsilon  = \frac{f_{e\tau}}{f_{\mu\tau}},  & &
\epsilon' = \frac{f_{e\mu}}{f_{\mu\tau}}.
\end{eqnarray}
With ${\cal M}^{\nu}.v_0=0$ one can express the parameters $\epsilon$ 
and $\epsilon'$ also in terms of the entries of the neutrino mass matrix. 
A straightforward calculation yields 
\begin{eqnarray}\label{epsmat}
\epsilon &=&\frac{m_{12}m_{33}-m_{13}m_{23}}{m_{22}m_{33}-m_{23}^2}, \\
\nonumber
\epsilon &=&\frac{m_{12}m_{23}-m_{13}m_{22}}{m_{22}m_{33}-m_{23}^2}.
\end{eqnarray}
Interestingly, eq. (\ref{epsmat}) can be rewritten directly as a function 
of the measured neutrino angles. For normal hierarchy, i.e. $m_{\nu_{1,2,3}} 
\simeq (0,m,M)$, one obtains
\footnote{We use the notation $m\simeq \sqrt{\Delta m^2_{\odot}}$ and 
$M\simeq \sqrt{\Delta m^2_{\rm Atm}}$, as well as $m_{\nu_3} \simeq 0$ for 
inverse hierarchy. This has the advantage that $\theta_{12}=\theta_{\odot}$,
$\theta_{23}=\theta_{\rm Atm}$ and $\theta_{13}=\theta_R$ for both 
hierarchies.}
\begin{eqnarray}\label{epsang}
\epsilon &=& \tan\theta_{12} \frac{\cos\theta_{23}}{\cos\theta_{13}}
+\tan\theta_{13}\sin\theta_{23} e^{-i\delta},\\ \nonumber
\epsilon' &=& \tan\theta_{12} \frac{\sin\theta_{23}}{\cos\theta_{13}}
- \tan\theta_{13}\cos\theta_{23} e^{-i\delta}.
\end{eqnarray}
Note, that eq. (\ref{epsang}) does not depend on neutrino masses,  
and that current data on neutrino angles require {\em both} $\epsilon$ 
and $\epsilon'$ to be non-zero. On the other hand, in the case of inverse 
hierarchy, $m_{\nu_{1,2,3}}\simeq (M,\pm M +m,0)$, eq. (\ref{epsmat}) leads 
to
\begin{eqnarray}\label{epsangI}
\epsilon &=& - \cot\theta_{13}\sin\theta_{23} e^{-i\delta},\\ \nonumber
\epsilon' &=& \cot\theta_{13}\cos\theta_{23} e^{-i\delta}.
\end{eqnarray}
Again, both $\epsilon$ and $\epsilon'$ have to be different from zero. 
Note that $\delta$ in eq. (\ref{epsang}) and (\ref{epsangI}) is a 
CP-violating phase, which reduces to a CP-sign $\delta = 0,\pi$ in case 
of real parameters.

With the equations outlined above, we are now in a position to give 
an estimate of the typical size of neutrino masses in the model. For 
an analytical understanding, the following approximation is quite 
helpful. Since $m_e \ll m_{\mu},m_{\tau}$, $\omega_{ee}$, $\omega_{e\mu}$ 
and $\omega_{e\tau}$ are expected to be much smaller than the other  
$\omega_{\alpha\beta}$. Then, in the limit $\omega_{ee} = \omega_{e\mu} = 
\omega_{e\tau}= 0$, eq. (\ref{mnu}) reduces to
\begin{equation}\label{mnusim}
{\cal M}^{\nu} = \zeta 
\left(\begin{array}{cccc}
\epsilon^2 \omega_{\tau\tau}+2 \epsilon \epsilon' \omega_{\mu\tau}+ 
\epsilon'^2\omega_{\mu\mu} & 
\epsilon \omega_{\tau\tau}+\epsilon' \omega_{\mu\tau} &
-\epsilon \omega_{\mu\tau} - \epsilon' \omega_{\mu\mu} \cr
& & \cr
\cdot &  \omega_{\tau\tau} &
 -\omega_{\mu\tau} \cr 
& & \cr
\cdot & \cdot &  \omega_{\mu\mu} 
\end{array}\right),
\end{equation}
where 
\begin{eqnarray}\label{defzeta}
\zeta & =& \frac{8\mu}{(16 \pi^2)^2}\frac{f_{\mu\tau}^2}{m_h^2}
{\cal I}(\frac{m_k^2}{m_h^2}).
\end{eqnarray}
From eq. (\ref{mnusim}) it is easy to estimate the typical ranges of 
parameters, for which the model can explain current neutrino data. In 
case of normal hierarchy, a large atmospheric angle requires $\omega_{\mu\mu} 
\simeq - \omega_{\mu\tau} \simeq \omega_{\tau\tau}$. Thus, we find 
the constraint
\begin{equation}\label{htyp}
h_{\tau\tau} \simeq (\frac{m_{\mu}}{m_{\tau}}) h_{\mu\tau} 
\simeq (\frac{m_{\mu}}{m_{\tau}})^2 h_{\mu\mu} .
\end{equation}
On the other hand, a solar angle of order $\tan\theta_{\odot}\simeq
\frac{1}{\sqrt{2}}$ requires $\epsilon \sim \epsilon' \simeq 1/2$, 
see eq.  (\ref{epsang}). Inverse hierarchy still requires $\omega_{\mu\mu} 
\simeq \omega_{\mu\tau} \simeq \omega_{\tau\tau}$, although with a different 
relative sign, while $\epsilon$ and $\epsilon'$ have to be much larger, i.e. 
$\epsilon \sim \epsilon' \simeq \frac{M}{m}$, see also eq. (\ref{epsangI}).

What is the maximal neutrino mass the model can generate? 
Using eqs (\ref{limmu}) and (\ref{htyp}), this upper limit can be 
estimated choosing $h_{\mu\mu}$ maximal. Motivated by perturbativity, 
we choose $h_{\mu\mu}=1$. \footnote{One could also choose 
$h_{\mu\mu}=\sqrt{4\pi}$ . However, as pointed out in \cite{Babu:2002uu}, 
even $h_{\mu\mu}=1$ at the weak scale will result in non-perturbative 
values of $h_{\mu\mu}$ at scales just one order of magnitude larger.}
Then, $m_{k} \gsim 800$ GeV is required (see the next section), and 
with $m_{h} = 100$ GeV, ${\cal I}(r) \simeq 0.3$ results. Putting finally 
$f_{\mu\tau} = 0.03$ we arrive at $m_{\nu_3} \simeq 0.05$ eV. Since all 
other parameters in this estimate have been put to extreme values, 
$f_{\mu\tau} \ge 0.03$ will be required in general. Obviously, even 
considering only neutrino data, the parameters of the model are already 
severely constrained.

\section{Flavour violating charged lepton decays}

Due to the flavour off-diagonal couplings of the $k^{++}$ and $h^{+}$ 
scalars to SM leptons, the model has sizeable non-standard flavour 
violating charged lepton decays. An extensive list of constraints on 
model parameters, derived from the observed upper limits of these decays, 
can be found in \cite{Babu:2002uu}. Here we will discuss decays of the 
type $l_{\alpha}\rightarrow l_{\beta} \gamma$ and their connection 
with neutrino physics. As the experimentally most interesting case 
we concentrate on $\mu\rightarrow e \gamma$. A short comment on $\tau$ 
decays is given at the end of this section.

\begin{figure}
\begin{center}
\includegraphics[width=45mm, height=30mm]{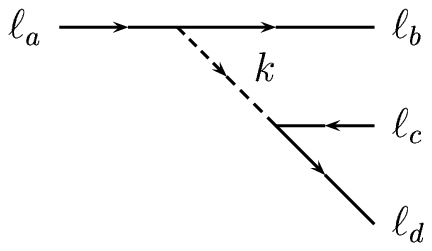}
\hskip20mm
\includegraphics[width=55mm, height=40mm]{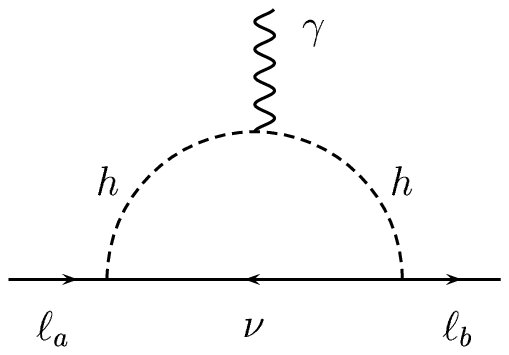}
\end{center}
\caption{Example diagrams for flavour changing charged lepton decays in 
the model. In addition to the diagrams shown, there 
are also box graphs involving $h^+$ contributing to $l_{a} \rightarrow 
l_{b}l_{c}l_{d}$, as well as graphs with $k^{++}$, similar to the one 
shown, contributing to $l_{a} \rightarrow l_{b}\gamma$.}
\label{fig:2lp}
\end{figure}

Consider the partial decay width of $l_{\alpha} \rightarrow l_{\beta}\gamma$ 
induced by the $h^+$ scalar loop shown in fig. (\ref{fig:2lp}). In the limit 
of $m_{\beta} \ll m_{\alpha}$ it is given by
\begin{equation}\label{eq:llgam}
\Gamma(l_{\alpha} \rightarrow l_{\beta}\gamma) = 2 \alpha 
m_{\alpha}^3(\frac{m_{\alpha}}{96\pi^2})^2
\Big(\frac{(f^{\dagger}f)_{\beta\alpha}}{m_h^2}\Big)^2.
\end{equation}
We will be interested in deriving a lower bound on the numerical 
value of eq. (\ref{eq:llgam}) in the following. Note, that although 
there is a graph similar to the one shown in fig. (\ref{fig:2lp}) with 
a $k^{++}$ in the intermediate state, there is no interference between 
the two contributions (in the limit where the smaller lepton mass is 
put to zero). Thus, in deriving the lowest possible value of 
Br($\mu\rightarrow e \gamma$) we will put the contribution from $k^{++}$ 
to zero. Any finite contribution from the doubly charged scalars would 
lead to stronger bounds on $m_h$ than the numbers quoted below. 

Using eqs (\ref{defeps}), (\ref{mnusim}) and (\ref{defzeta}) we can 
rewrite eq. (\ref{eq:llgam}) as 
\begin{eqnarray}\label{eq:muegam}
Br(\mu \rightarrow e\gamma) & = &
\frac{\alpha\epsilon^2 m_{\mu} \pi^3}{18 \sqrt{6}\Gamma_{\mu} 
h_{\mu\mu}^2 {\cal I}(r)^2 }  \frac{m_{\nu}^2}{m_h^2}\\ 
\label{eq:muegam_num}
& \simeq & 4.5 \cdot 10^{-10} 
\Big(\frac{\epsilon^2}{h_{\mu\mu}^2{\cal I}(r)^2}\Big) 
\Big(\frac{m_{\nu}}{\rm 0.05 \hskip1mm eV}\Big)^2
\Big(\frac{\rm 100 \hskip1mm GeV}{m_{h}}\Big)^2
\end{eqnarray}
With $\epsilon$ non-zero, constrained by eq. (\ref{epsang}) or eq. 
(\ref{epsangI}) in case of normal or inverse hierarchy, 
Br($\mu \rightarrow e\gamma$) has to be non-zero as well. Its smallest 
numerical value is found for the largest possible value of $h_{\mu\mu}$ 
and ${\cal I}(r)$. 

\begin{figure}
\centering
\includegraphics[width=6.6cm, height=6cm]{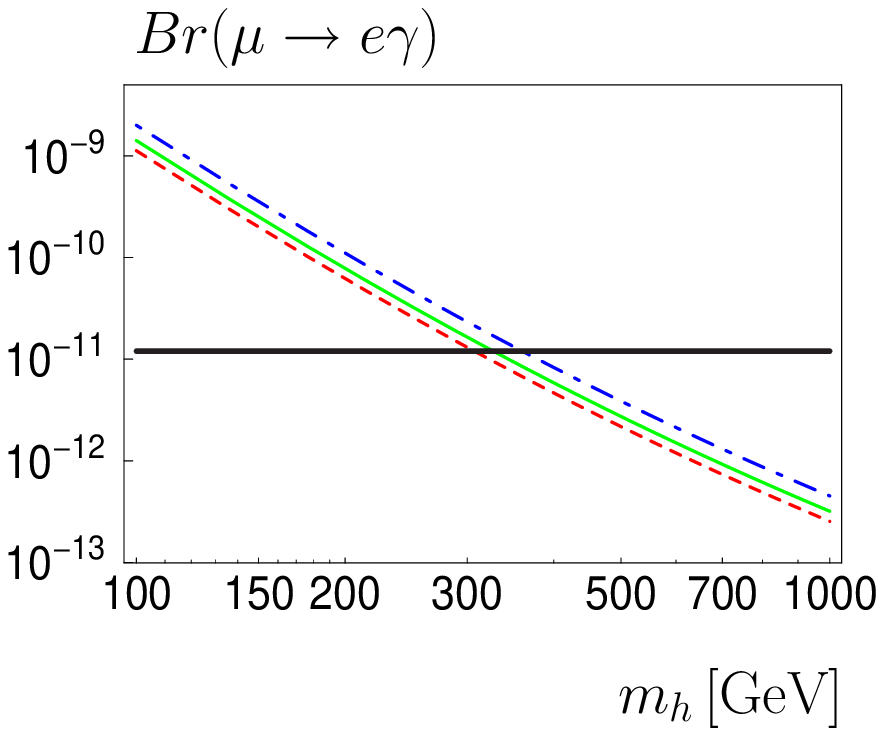}
\hspace{0.2cm}
\includegraphics[width=6.6cm, height=6cm]{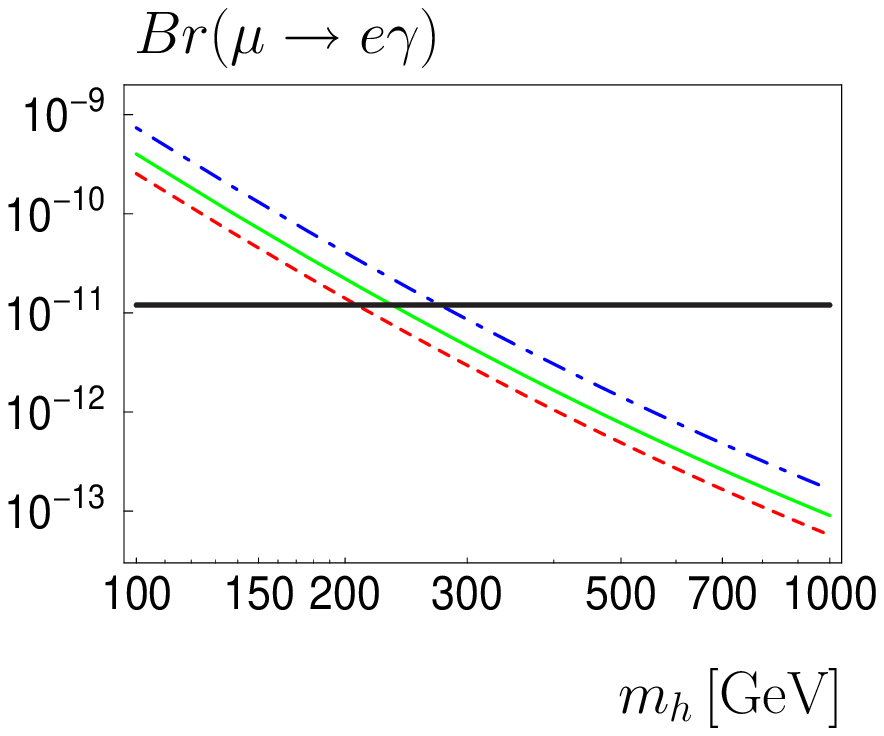}
\caption{Conservative lower limit on the branching ratio 
Br($\mu\rightarrow e\gamma$) as a function of the charged scalar 
mass $m_h$ for normal hierarchy. The three lines are for the current 
solar angle $\sin^2\theta_{12}$ best fit value (full line) and 3 $\sigma$ 
lower (dashed line) and upper (dot-dashed line) bounds. To the left 
$\delta =0$, to the right $\delta =\pi$. Other parameters fixed at 
$\sin^2\theta_{23}=0.5$, $\sin^2\theta_{13}=0.040$ and 
$\Delta m^2_{\rm Atm} = 2.0 \cdot 10^{-3}$ eV$^2$.}
\label{MuEGamVmH}
\end{figure}

In order to calculate ${\cal I}(r)$ we need to fix $r$ consistent with 
all experimental constraints. This is done in the following way.
The decay width $l_{a} \rightarrow l_{b}l_{c}l_{d}$ induced by virtual 
exchange of $k^{++}$, see fig. (\ref{fig:2lp}), is, in the limit 
$m_b,m_c,m_d \ll m_a$,
\begin{equation}\label{eq:3lep}
\Gamma(l_{a} \rightarrow l_{b}l_{c}l_{d}) = 
\frac{1}{8}\frac{m_a^5}{192\pi^3}\Big|\frac{h_{ab}h_{cd}^*}{m_k^2}\Big|^2.
\end{equation}
The most relevant constraint for the current discussion is derived from 
the upper bound on $\tau \rightarrow 3 \mu$ decay, which yields,
\begin{equation}\label{t3m}
\frac{|h_{\mu\tau}h_{\mu\mu}|}{m_{k}^2} \lsim 10^{-7} 
\hskip2mm {\rm GeV}^{-2}.
\end{equation}
For $h_{\mu\tau}(\frac{m_{\tau}}{m_{\mu}}) = h_{\mu\mu}= 1$, this bound 
implies $m_{k} \gsim 770$ GeV. For any fixed value of $h_{\mu\mu}$, 
we can therefore fix the minimum value of $r$, i.e. the maximum allowed 
value of ${\cal I}(r)$, which in turn fixes the lower bound on 
Br($\mu\rightarrow e\gamma$). 

Fig. (\ref{MuEGamVmH}) shows the resulting lower limit on 
Br($\mu\rightarrow e\gamma$) as a function of the charged scalar 
mass $m_h$ for the case of normal hierarchy. In this plot, we have assumed 
that the parameters $\mu$, $h_{\mu\mu}$ (and $\Delta m^2_{\rm Atm}$) take 
their maximal (minimal) allowed values, thus we consider this limit 
conservative. We would like to stress again, that any non-zero 
contributions to the decay $\mu\rightarrow e\gamma$ from $k^{++}$ can 
only increase Br($\mu\rightarrow e\gamma$). 

\begin{figure}
\centering
\includegraphics[width=6.6cm, height=6.1cm]{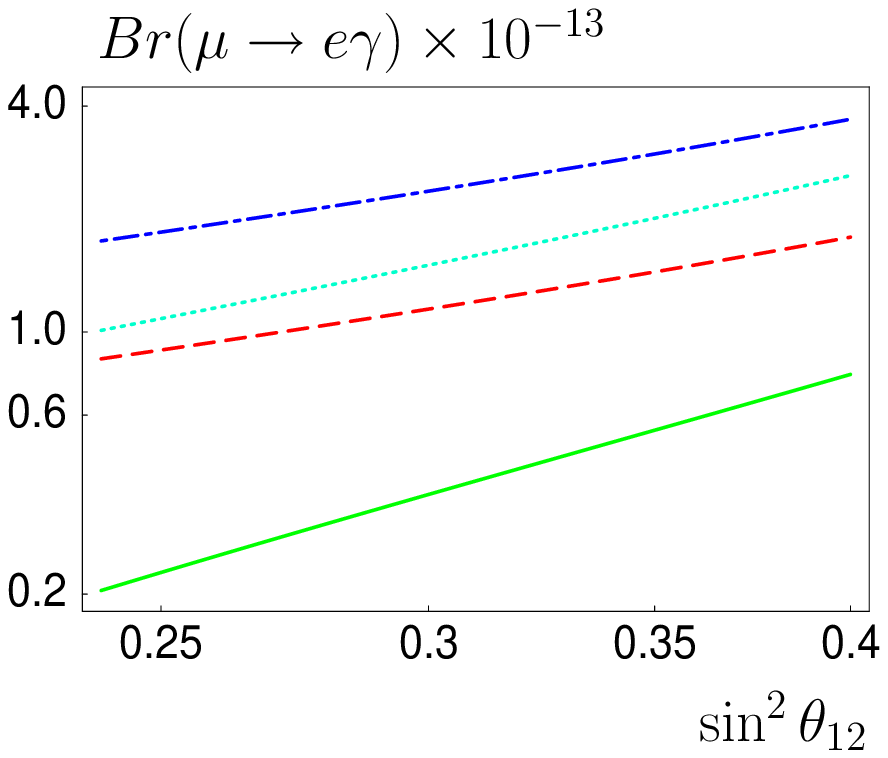}
\hspace{0.2cm}
\includegraphics[width=6.6cm, height=6.1cm]{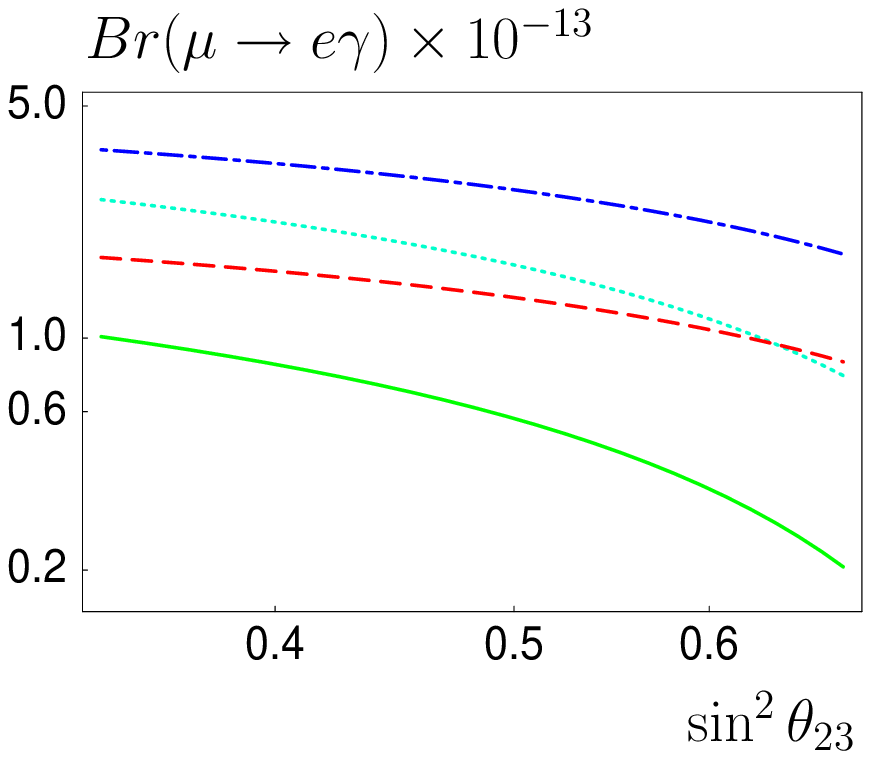}
  
\caption{Dependence of the lower limit on Br$(\mu\rightarrow e\gamma)$ 
for normal hierarchy on neutrino angles, for $h_{\mu\mu}=1,\delta=\pi,
m_h=10^3\mbox{GeV}$. Left plot: ($\sin^2\theta_{23}=0.68$, 
$\sin^2\theta_{13}=0$) dashed line, ($\sin^2\theta_{23}=0.68$, 
$\sin^2\theta_{13}=0.040$) full line, ($\sin^2\theta_{23}=0.34$, 
$\sin^2\theta_{13}=0$) dash-dotted line, ($\sin^2\theta_{23}=0.34$, 
$\sin^2\theta_{13}=0.040$) dotted line. Right plot: 
($\sin^2\theta_{12}=0.40$, $\sin^2\theta_{13}=0$) dash-dotted line, 
($\sin^2\theta_{12}=0.40$, $\sin^2\theta_{13}=0.040$) dotted line, 
($\sin^2\theta_{12}=0.24$, $\sin^2\theta_{13}=0$) dashed line, 
($\sin^2\theta_{12}=0.24$, $\sin^2\theta_{13}=0.040$) full line.}
\label{MuEGamVAng}
\end{figure}

\begin{figure}
\centering
\includegraphics[width=6.6cm, height=6.1cm]{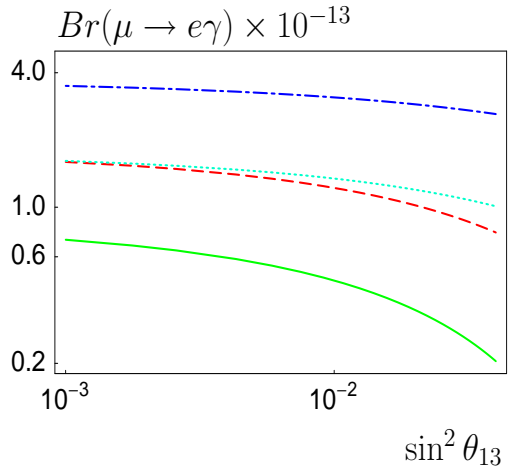}
\caption{Dependence of the lower limit on Br$(\mu\rightarrow e\gamma)$ 
for normal hierarchy on the reactor angle, for $h_{\mu\mu}=1,\delta=\pi,
m_h=10^3\mbox{GeV}$. Other parameters are chosen as ($\sin^2\theta_{23}=0.68$,
$\sin^2\theta_{12}=0.40$) dashed line, ($\sin^2\theta_{23}=0.68$,
$\sin^2\theta_{12}=0.24$) full line, ($\sin^2\theta_{23}=0.34$,
$\sin^2\theta_{12}=0.40$) dash-dotted line and ($\sin^2\theta_{23}=0.34$, 
$\sin^2\theta_{12}=0.24$) dotted line.}
\label{MuEGamVthR}
\end{figure}

Fig. (\ref{MuEGamVAng}) and (\ref{MuEGamVthR}) show the dependence of the 
limit on Br($\mu\rightarrow e\gamma$) on the three neutrino angles. Both 
plots are for the case of normal hierarchy. Larger values of $\theta_{12}$ 
($\theta_{23}$) result in larger (smaller) upper bounds. Smaller ranges 
of these parameters obviously lead to tighter predictions. For $\theta_{13}$, 
below approximately $\sin^2\theta_{13} \lsim 0.01$ the dependence of 
Br($\mu\rightarrow e\gamma$) is rather weak. 

%
\begin{figure}[t]
  \centering
  \includegraphics[width=6.6cm, height=6.1cm]{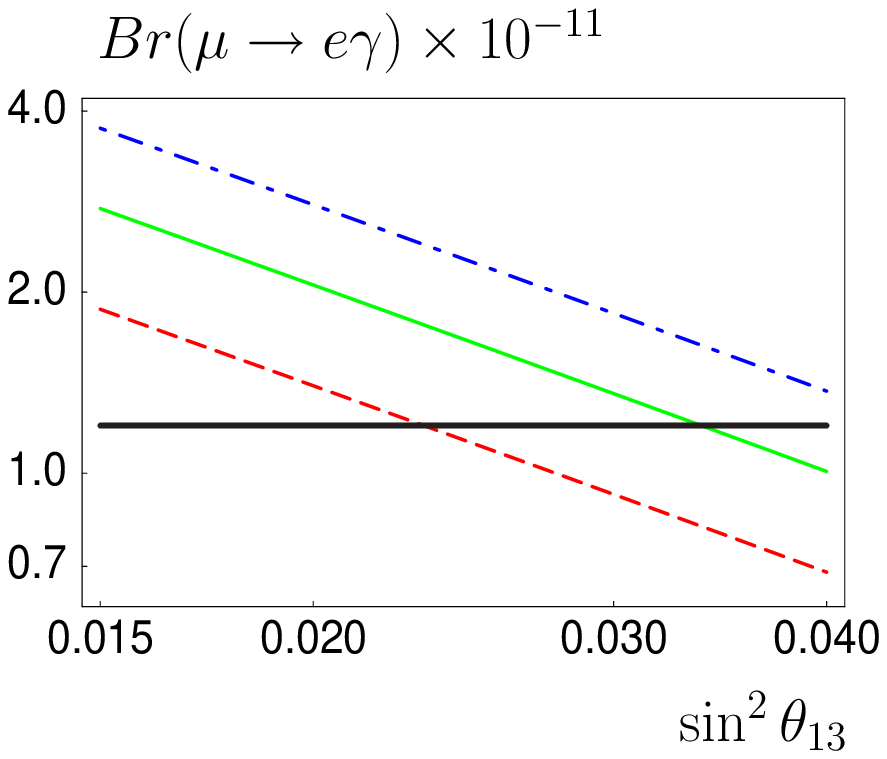}
\hskip2mm
  \includegraphics[width=6.6cm, height=6.1cm]{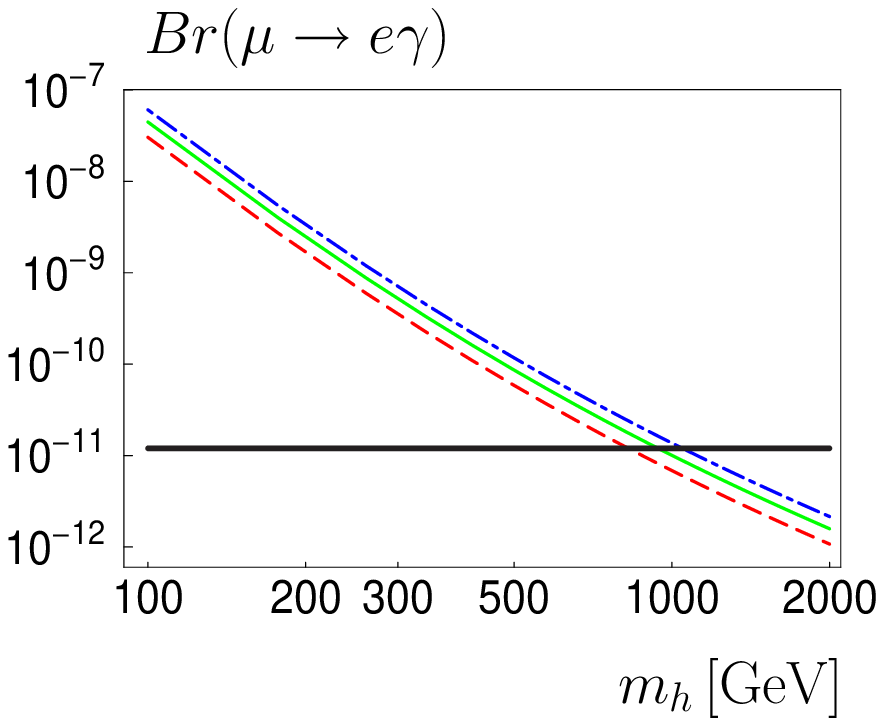}

\caption{Lower limit on Br($\mu \rightarrow e \gamma$) for inverted 
hierarchy, to the left: versus the reactor angle; to the right: 
versus $m_h$. Parameter choices as before. 
The three lines are for the current $\sin^2\theta_{23}$ best fit 
value (full line) and 3 $\sigma$ upper (dot-dashed line) and lower
(dashed line) bounds.}
\label{MuEGamInv}
\end{figure}

Fig. (\ref{MuEGamInv}) shows the calculated lower limit on 
Br($\mu \rightarrow e \gamma$) for the case of inverted hierarchy, both, 
versus the reactor angle and versus $m_h$. Due to the fact that $\epsilon$ 
must be larger than $\epsilon \simeq \sin\theta_{23}/\tan\theta_{13}$, 
the expected values for Br($\mu \rightarrow e \gamma$) turn out to be 
much bigger than for the case of normal hierarchy. Even 
Br($\mu \rightarrow e \gamma$) $\lsim 10^{-11}$ requires already 
TeV-ish masses for $m_h$.

The most conservative limits for $m_h$ are always found for $\delta=\pi$, 
$\sin^2\theta_{12}=(\sin^2\theta_{\odot})^{Min}$, 
$\sin^2\theta_{23}=(\sin^2\theta_{\rm Atm})^{Max}$ 
and $\sin^2\theta_{13}=(\sin^2\theta_{\rm R})^{Max}$. 
For the current bound of $Br(\mu\to e\gamma)\le 1.2\times10^{-11}$, 
we find $m_{h}\ge 160\,\mbox{GeV}$ ($m_h=825\,\mbox{GeV}$) for 
normal (inverse) hierarchy. Future experiments \cite{exp:meg} 
expect to lower this limit to $Br(\mu\to e\gamma)\le 10^{-13}$, 
resulting in $m_{h}\ge 590\,\mbox{GeV}$ ($m_h=5040 \,\mbox{GeV}$). 
Given these numbers, one expects that the MEG experiment \cite{exp:meg} 
will see the first evidence for $\mu \rightarrow e \gamma$ in the 
near future, if the model discussed here indeed is the origin of 
neutrino masses.

Finally, we would like to mention that the decays $\tau\rightarrow \mu\gamma$ 
and $\tau\rightarrow e\gamma$ can be constrained in a similar way. 
However, the resulting lower limits, also of order ${\cal O}(10^{-13})$, 
are far below the near-future experimental sensitivities and thus 
less interesting.

\section{Accelerator tests of the model}

In this section we will briefly discuss some possible accelerator signals 
of the model. With the couplings of $h^+$ and $k^{++}$ tightly constrained 
by neutrino physics and flavour violating lepton decays, it turns out 
that various decay branching ratios can be predicted. Observing the 
corresponding final states could serve as a definite test of the model 
as the origin of neutrino masses. 

In \cite{Babu:2002uu} it has been estimated that at the LHC discovery of 
$k^{++}$ might be possible up to masses of $m_{k}\le 1$ TeV approximately. 
In the following we will therefore always assume that $m_{k}\le 1$ TeV and, 
in addition, $m_h \le 0.5$ TeV. Given the discussion of the previous 
section, this range of masses implies that $\mu \rightarrow e \gamma$ 
should be seen at the MEG experiment.

\noindent
The $h^{+}$ will decay to leptons with a partial decay width of, in the 
limit $m_{\alpha}=0$,
\begin{equation}\label{hdec}
\Gamma(h^{+} \rightarrow l_{\alpha}\sum_{\beta}{\nu_\beta}) = 
\frac{m_{h}}{16 \pi}\sum_{\beta} f_{\alpha\beta}^2.
\end{equation}

\noindent
We can re-express eq. (\ref{hdec}) in terms of the parameters $\epsilon$ 
and $\epsilon'$ as
\begin{eqnarray}\label{hbr}
Br(h^{+} \rightarrow e\sum_{\beta}{\nu_\beta}) = 
\frac{\epsilon^2 + \epsilon'^2}{2(1 + \epsilon^2 + \epsilon'^2)}, \\
\nonumber
Br(h^{+} \rightarrow \mu\sum_{\beta}{\nu_\beta}) = 
\frac{1 + \epsilon'^2}{2(1 + \epsilon^2 + \epsilon'^2)}, \\
\nonumber
Br(h^{+} \rightarrow \tau\sum_{\beta}{\nu_\beta}) = 
\frac{1 + \epsilon^2 }{2(1 + \epsilon^2 + \epsilon'^2)} .
\end{eqnarray}

\begin{figure}[ht]
\begin{center}
\includegraphics[width=6.0cm, height=5.4cm]{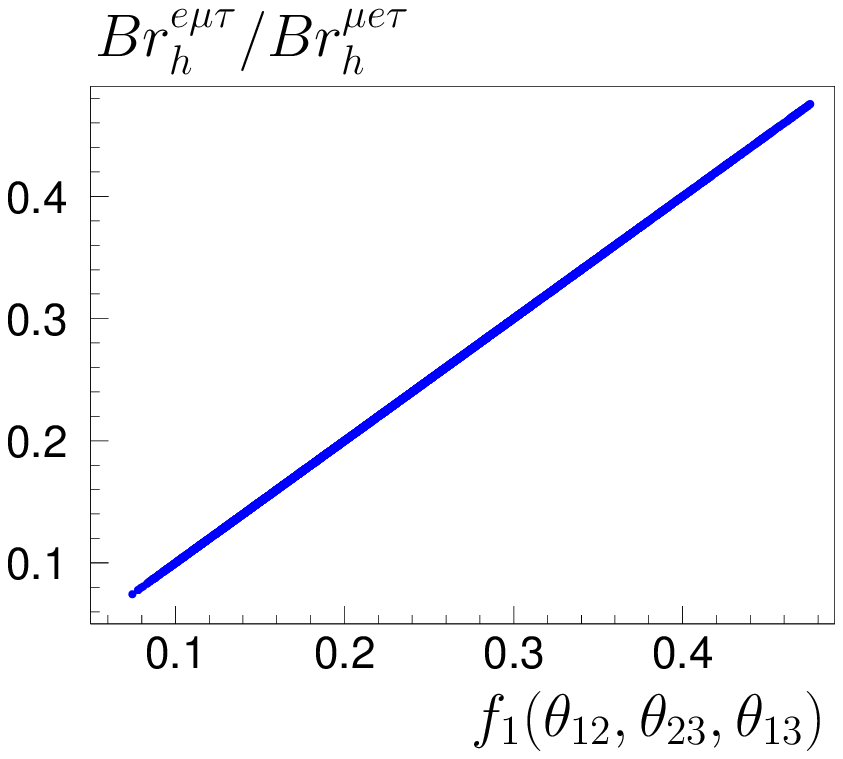}
\hspace{0.5cm}
\includegraphics[width=6.0cm, height=5.4cm]{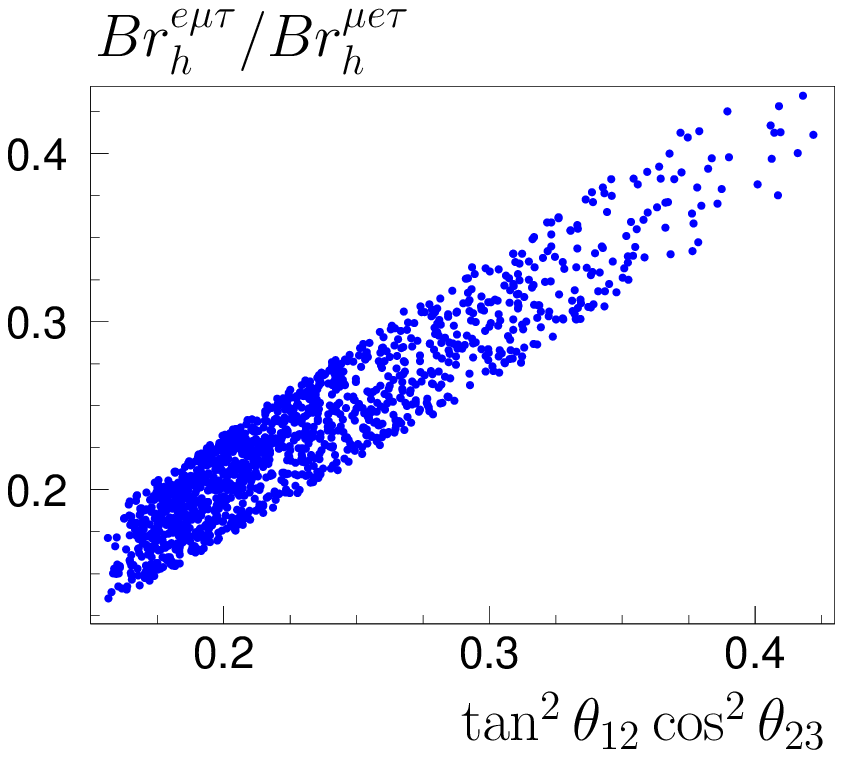}
\end{center}
\begin{center}
\includegraphics[width=6.0cm, height=5.4cm]{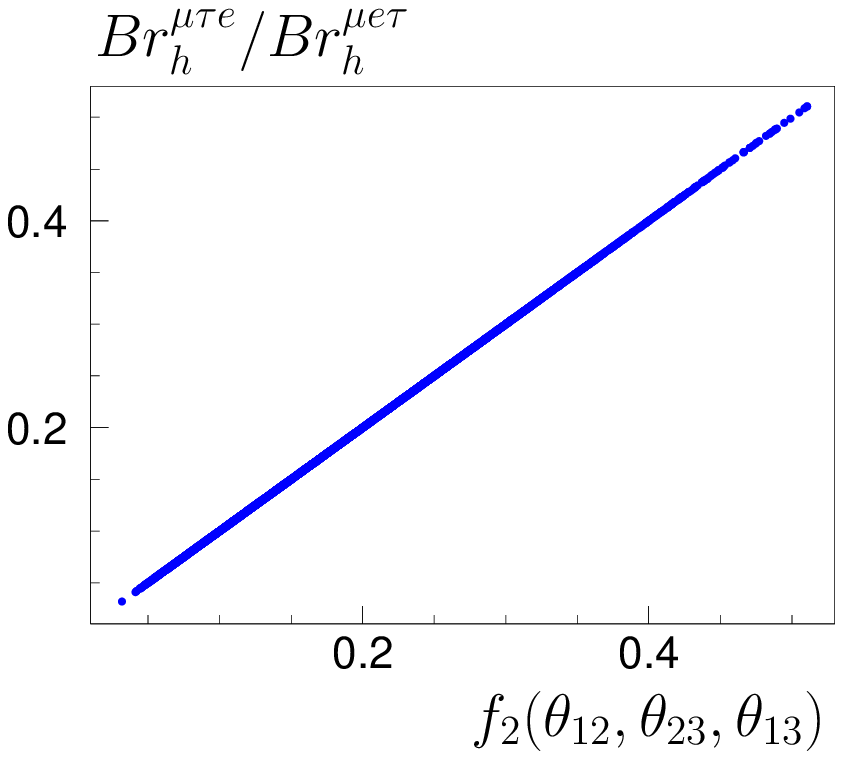}
\hspace{0.5cm}
\includegraphics[width=6.0cm, height=5.4cm]{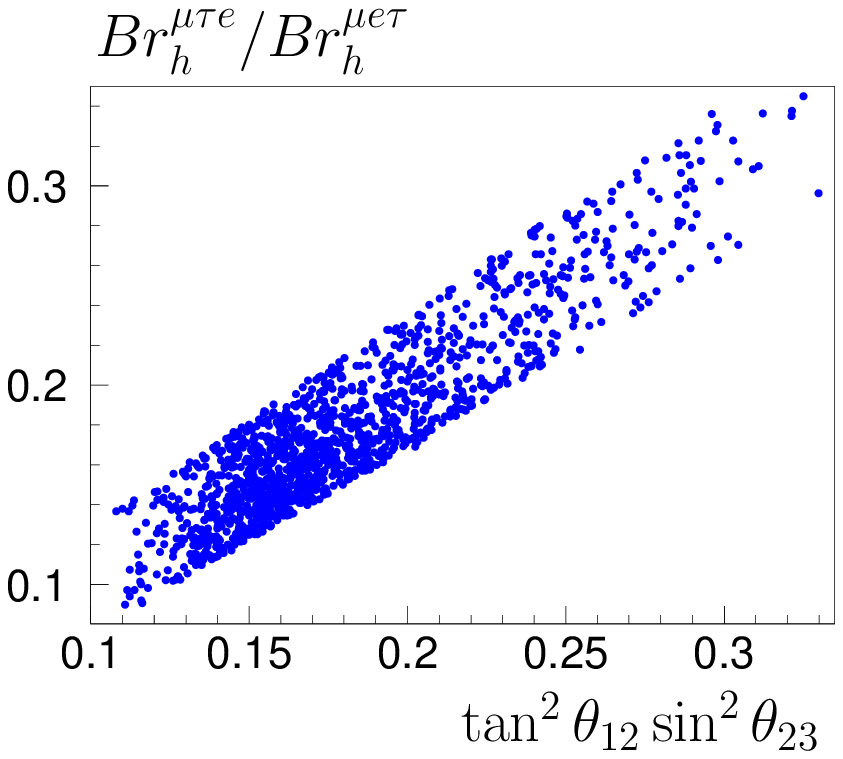}
\end{center}
\caption{Ratios of decay branching ratios $Br_h^{ijk}$, see text, versus 
$f_1(\theta_{12},\theta_{23},\theta_{13})= \tan\theta_{12}
\frac{\cos\theta_{23}}{\cos\theta_{13}}+\tan\theta_{13}\sin\theta_{23}$ 
(top); and 
$f_2(\theta_{12},\theta_{23},\theta_{13})=\tan\theta_{12}
\frac{\sin\theta_{23}}{\cos\theta_{13}}-\tan\theta_{13}\cos\theta_{23}$ 
(bottom). In the right plots $\sin^2\theta_{13}<2.5\times 10^{-3}$ has 
been assumed. All points satisfy updated experimental neutrino data.}
\label{fig:brijk}
\end{figure}

\noindent
It is therefore possible to directly ``measure'' $\epsilon^2$ or $\epsilon'^2$ 
by calculating ratios of branching ratio differences, such as the ones 
shown in fig. (\ref{fig:brijk}). Here, 
\begin{equation}\label{def:brijk}
Br_h^{ijk}\equiv Br(h^-\to\nu\ell_i^-)-
    Br(h^-\to\nu\ell_j^-)+Br(h^-\to\nu\ell_k^-).
\end{equation}
The plots on the left in fig. (\ref{fig:brijk}) show calculated 
ratios of branching ratios versus eq. (\ref{epsang}), i.e. normal 
hierarchy, versus $\epsilon$ (top) and $\epsilon'$ (bottom). All 
points are obtained by numerically diagonalizing eq. (\ref{mnu}) 
for random parameters and checking for consistency with all experimental 
constraints. However, since $\theta_{13}$ is unkown, eq. (\ref{epsang}) 
can be numerically calculated, but at the moment not experimentally 
determined. Thus, the plots on the right of the figure show the same 
ratios of branching ratios, but versus $(\tan\theta_{12}\cos\theta_{12})^2$ 
and $(\tan\theta_{12}\sin\theta_{12})^2$. The cut on $\sin^2\theta_{13}$ 
of $\sin^2\theta_{13}<2.5\times 10^{-3}$ in this plot is motivated by 
the expected sensitivity of the next generation of reactor experiments 
\cite{Ardellier:2004ui,Huber:2006vr}. The width of the band of points 
in these plots indicates the uncertainty with which the corresponding 
ratios can be predicted. 

In case of normal (inverse) hierarchy, assuming best fit parameters for 
the neutrino angles, eq. (\ref{hbr}) indicates that the branching ratios 
for $e,\mu$ and $\tau$ final states of $h^+$ decays should scale as 
$2/12 : 5/12 : 5/12$ ($1/2:1/4:1/4$). Inserting the current 3 $\sigma$ 
ranges of the angles, following eqs. (\ref{epsang}) and (\ref{epsangI}) 
results in the following predicted ranges
\begin{eqnarray}\label{hbrmaxmin}
Br(h^{+} \rightarrow e\sum_{\beta}{\nu_\beta}) = 
[0.13,0.22] \hskip5mm ([0.48,0.50]) \\ \nonumber
Br(h^{+} \rightarrow \mu\sum_{\beta}{\nu_\beta}) = 
[0.31,0.50]  \hskip5mm ([0.17,0.34])\\ \nonumber
Br(h^{+} \rightarrow \tau\sum_{\beta}{\nu_\beta}) = 
[0.31.0.50]  \hskip5mm  ([0.18,0.35])
\end{eqnarray}
for normal (inverse) hierarchy. The different predicted branching ratios 
for final states with electrons should make it nearly straightforward to 
distinguish normal and inverse hierarchy. Measuring any branching ratio 
outside the range given in eq. (\ref{hbrmaxmin}) would rule out the 
model as possible origin of neutrino masses.

The doubly charged scalar of the model decays either to two same-sign 
leptons or to two $h^+$ final states. The partial width to leptons is, 
for $m_{\alpha},m_{\beta}=0$, 
\begin{equation}\label{kdeclep}
\Gamma(k^{++} \rightarrow l_{\alpha}l_{\beta}) = 
\frac{m_{k}}{16 \pi} h_{\alpha\beta}^2
\end{equation}
whereas the decay width to two $h^{+}$ can be calculated to be
\begin{equation}\label{kdech}
\Gamma(k^{++} \rightarrow h^{+}h^{+}) = 
\frac{1}{16 \pi} \frac{\mu^2}{m_{k}}
\beta(\frac{m_{h}^2}{m_{k}^2})
\end{equation}
Here, $\beta(x^2)=\sqrt{1-4 x^2}$ is a kinematical factor. 

The couplings $h_{\alpha\beta}$ in eq.(\ref{kdeclep}) are constrained by 
neutrino physics, see eq.(\ref{htyp}), and by lepton flavour violating 
decays of the type $l_a\rightarrow l_bl_cl_d$. For $m_k \le 1$ TeV the 
couplings $h_{ee}$, $h_{e\mu}$ and $h_{e\tau}$ are constrained to be 
smaller than $0.4$, $4\cdot 10^{-3}$ and $7\cdot 10^{-2}$ \cite{Babu:2002uu}. 
Thus, the leptonic final states of $k^{++}$ decays are mainly like-sign 
muon pairs (and possibly electrons).

\begin{figure}
\centering
\includegraphics[width=6.6cm, height=6cm]{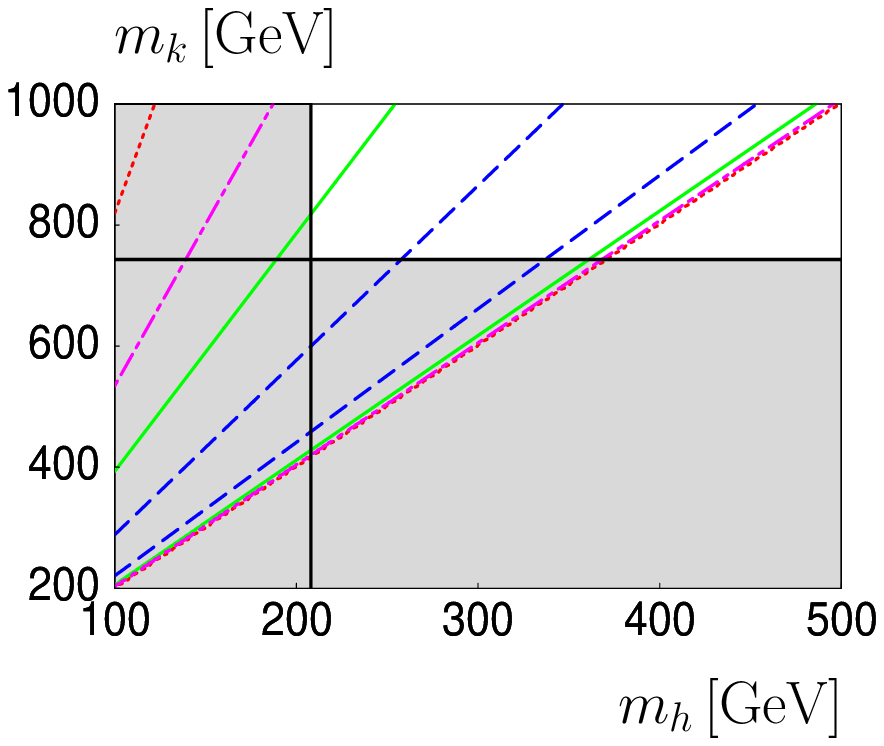}
\hspace{0.2cm}
\includegraphics[width=6.6cm, height=6cm]{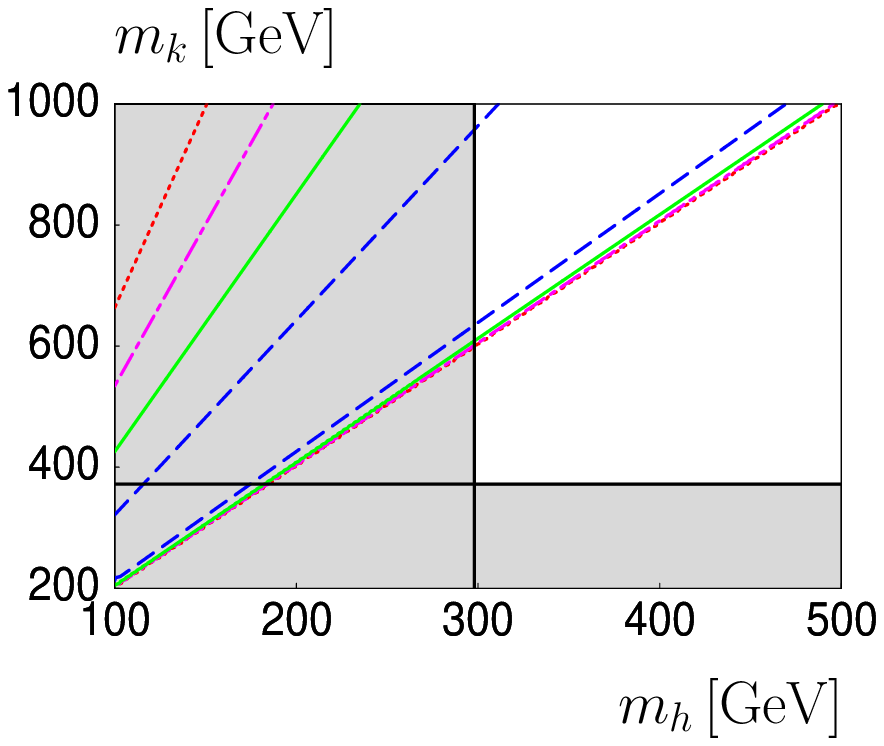}
\caption{Lines of constant Br$(k^{++} \rightarrow h^{+}h^{+})$, 
assuming to the left $h_{\mu\mu}=1$: $Br^{hh}_k=0.1,0.2,0.3$ and $0.4$ 
for dotted, dash-dotted, full and dashed line. The vertical line 
corresponds to $m_h=208\,\mbox{GeV}$ for which $Br(\mu\to e\gamma)=
1.2\times10^{-11}$ and horizontal line to $m_k=743\,\mbox{GeV}$ for 
which $Br(\tau\to 3\mu)=1.9\times10^{-6}$, i.e. parameter combinations 
to the left/below this line are forbidden. Plot on the right assumes 
$h_{\mu\mu}=0.5$. Lines are for $Br^{hh}_k=0.4,0.5,0.6$ and $0.7$, 
dotted, dash-dotted, full and dashed line. Again the shaded regions 
are excluded by Br$(\mu\to e\gamma)$ and Br$(\tau\to 3\mu)$.}
\label{fig:brkhh}
\end{figure}

An interesting situation arises, if $m_k \ge 2 m_h$. In this case, one 
can measure the lepton number violating parameter $\mu$ of eq.(\ref{scalar}) 
by measuring the branching ratio of $k^{++} \rightarrow h^{+}h^{+}$. 
Combining eq. (\ref{kdeclep}) and eq. (\ref{kdech}) we can write
\begin{equation}\label{brkdech}
Br(k^{++} \rightarrow h^{+}h^{+}) \simeq 
\frac{\mu^2\beta}{m_{k}^2h_{\mu\mu}^2+\mu^2\beta}
\simeq 
\frac{f m_{h}^2\beta}{m_{k}^2h_{\mu\mu}^2+f m_{h}^2\beta}.
\end{equation}
Here, $h_{ee}\ll h_{\mu\mu}$ has been assumed. (For non-zero $h_{ee}$ 
replace simply $h_{\mu\mu} \rightarrow h_{\mu\mu} + h_{ee}$ in eq. 
(\ref{brkdech}).) Plots of constant Br$(k^{++} \rightarrow h^{+}h^{+})$ 
in the plane ($m_k,m_h$) are shown in fig. (\ref{fig:brkhh}).
Here, $\mu = f m_h$, with $f= (6\pi^2)^{1/4}$ has been used. 

Fig. (\ref{fig:brkhh}) shows the resulting branching ratios for 2 values 
of $h_{\mu\mu}$, fixing in both cases the couplings $f_{\alpha\beta}$ such 
that the atmospheric neutrino mass is correctly reproduced. For 
$h_{\mu\mu}\lsim 0.2$ the current limit on Br$(\mu\to e\gamma)$ rules 
out all $m_h \lsim 0.5$ TeV, thus this measurement is possible only 
for $h_{\mu\mu}\gsim 0.2$. Note that smaller values of $\mu$ lead to 
smaller neutrino masses, thus upper bounds on the branching ratio for 
$Br^{hh}_k$ can be interpreted as upper limit on the neutrino mass 
in this model.

\section{Conclusion}

The observed smallness of neutrino masses could be understood if it has  
a radiative origin. In this paper, we have studied some phenomenological 
aspects of one well-known incarnation of this idea 
\cite{Zee:1985id,Babu:1988ki}, in which neutrino masses arise only at 
2-loop order. 

Given the observed neutrino masses and angles, it turns out that the 
parameters of this model are very tightly constrained already today 
and thus it is possible to make various predictions for the near future. 
Perhaps the phenomenologically most important one is, that one expects 
that the process $\mu \rightarrow e\gamma$ has to be observed in the 
next round of experiments, i.e. Br($\mu \rightarrow e\gamma$) $\ge 10^{-13}$ 
is guaranteed for singly charged scalar masses smaller than 590 GeV 
(5.04 TeV) for normal (inverse) hierarchical neutrino masses, and 
larger or even much larger branching ratios are expected in general. 
At least for the case of inverse hierarchy an upper limit on the decay 
$\mu \rightarrow e\gamma$ of this order would certainly remove most of 
the motivation to study this model. 

On the other hand, if $\mu \rightarrow e\gamma$ is observed in the 
near future, it will be interesting to search for the charged scalars 
of the model at the LHC. As we have shown, in this case, several branching 
ratios of the decays of both, the singly and the doubly charged scalar 
are tightly fixed, mainly by data on neutrino angles. Observation of 
branching ratios outside the ranges discussed, would then definitely 
rule out the model as a possible explanation of neutrino masses. 

\section*{Acknowledgments}
This work was supported by Spanish grant FPA2005-01269, by the European 
Commission Human Potential Program RTN network MRTN-CT-2004-503369.  
M.H. is supported by a MCyT Ramon y Cajal contract. D.A.S. is supported 
by a Spanish PhD fellowship by M.C.Y.T.

\end{document}